# Effect of lattice relaxation on electronic spectra of helically twisted trilayer graphene: Large-scale atomistic simulation approach

Joonho Jang

Department of Physics and Astronomy and Institute of Applied Physics, Seoul National University, Seoul, Korea

Twisted trilayer graphene hosts two moiré superlattices originating from two interfaces between graphene layers. However, the system is generally unstable to lattice relaxation at small twist angles and is expected to show a significantly modified electronic band structure. In particular, a helical trilayer graphene - whose two twisted angles have the same sign - provides an attractive platform with a flat band isolated by large energy gaps near the magic angle, but the interplay between the lattice and the electronic degrees of freedom is not well understood. Here, we performed a large-scale molecular dynamics simulation to study the lattice relaxation of helical trilayer graphenes and evaluated their electronic spectra with a tight-binding model calculation. The comparison of the electronic spectra both with and without the lattice relaxation reveals how the lattice relaxation significantly modifies the electronic spectra particularly near the charge neutrality point. We also investigated the local density of states to visualize the spatially-varying electronic spectra that accords with domain patterns of moiré lattice stackings. We propose these characteristic spectral features in the electronic degrees of freedom of a relaxed helical trilayer graphene to be confirmed by scanning probe techniques, such as scanning single-electron transistors and scanning tunneling microscopes.

Email: Joonho.jang@snu.ac.kr

# I. INTRODUCTION

Twisted multilayer graphenes, where graphene layers are stacked with crystallographic axes of individual layers artificially twisted, provide a new opportunity to dramatically tune the electronic band structure [1,2] and realize emergent quantum phases [3–6]. In particular, the emergence of nearly flat electronic bands near the charge neutrality point (CNP) in "magic-angle" twisted bilayers was one of the milestones in the direction towards engineering quantum interaction and nontrivial band topology in these systems, leading to the observation of exotic quantum phases, such as correlated insulators and superconductivity [7,8]. The power of this approach is that it is extendable to other 2D materials systems such as transition metal dichalcogenides and oxide materials, which enriches the twistronics platform and opens a way to the realization of robust fractional quantum Hall effects and unconventional superconducting phases [9,10].

The effect of twisting layers on the electronic spectra of twisted multilayer systems, when the lattice relaxation is negligible or in the rigid lattice limit, can be calculated fairly straightforward via continuum-limit calculations and tight-binding methods [11,12]; however, properly accounting for the lattice relaxation is subtle and sometimes becomes a complex subject. For example, in twisted bilayer graphene, it was found that twisting with a small angle below ~ 0.7 deg results in alternating domains of a fully relaxed sublattice-locked system due to lattice relaxation while the lattice relaxation affects the electronic spectra in a lesser degree in angles above ~ 1.0 deg [13,14]. This can be qualitatively understood by considering the fact that the size of the superlattice decreases as the twisting angle increases but the amount of lattice coordinates that can be relaxed doesn't change significantly. Thus, near the magic angle at ~1.1 deg, the relaxation effect is not strong and considered to modify the electron bands only marginally.

In the case of trilayers, where two interfacial moiré structures coexist, an additional super-structure is expected to emerge with a larger spatial extent so that even small relaxation can affect the electronic spectra significantly. Among those, the helical trilayer graphene (HTTG) system can realize a higher order generalization of twisted bilayer graphene systems; HTTG has two moiré lattices - one from the interface between the first and second layers, and the other one from the interface between the second and third layers - whose axes are again twisted. In this case, the higher-order superlattice structure out of the twisted moiré bilayer develops with the length scale of a few hundreds nm when twisted by a few degrees. This so-called "super-moiré" or "moiré-of-moiré" was discussed in various contexts [15,16]. Without any lattice relaxation, a quasicrystal physics is expected in principle [17]; however, the lattice relaxation is found to be critical and low energy electronic states can become highly non-trivial [18,19]. Because layers are twisted with the same angles consecutively in HTTG, the two resulting moiré superlattices have the identical lattice constant and thus are generally susceptible for lattice relaxation that can align the two moiré superlattices commensurately. Recently, Devakul et al. [20] and Najatsuju et al. [21] discussed the effects of the lattice relaxation and the formation of macroscopic domains when the twisting angle is as large as a few degrees, with calculations based on configurational space and effective continuum-based models [13,15]. However, the trilayer and multilayer systems generally don't have a commensurate unit cell that limits the computational phase space and thus a realistic and accurate calculation of relaxation is challenging. For studies of electronic phases in HTTG and multilayer graphene systems, a more accessible, flexible and realistic method to simulate relaxation and investigate resulting moiré and supermoiré structures thus would be much desired.

In this paper, we performed real-space molecular dynamics (MD) simulations of HTTG in realistically-large size samples in conjunction with a tight-binding calculation to investigate the lattice relaxation and evaluate the modified electronic spectra. Our work presents a straight-forward and systematic approach to study HTTG for the subtle interplay between the lattice and electronic degrees of freedom and this method would be easily extendable to other incommensurate Van der Waals multilayer systems.

## II. Methods

### II.1. Molecular dynamics simulation

We chose the Large-scale Atomic/Molecular Massively Parallel Simulator (LAMMPS) [22], an open-source package for multipurpose molecular dynamics simulations, that supports efficient parallel computation schemes for systems with a large number of constituent particles. The lattice relaxation of multilayer graphenes can be easily implemented in this framework, which accommodates large-size layers of atoms and provides powerful and semi-accurate options for the Carbon-to-Carbon interatomic potentials that properly describe multilayer graphene samples.

To alleviate the consequence of the lack of a periodic structure and the absence of an exact unit cell, we used samples whose sizes were sufficiently large so that multiple unit-cells can be accommodated and sample edge effects become negligible. We start with three identical intrinsic graphene layers with lateral size ranging from *500 nm* to *1000 nm*, and the initial coordinates of Carbon atoms in the graphene layers are calculated based on the graphene's intrinsic hexagonal lattice with the lattice constant *$a = 0.246\ nm$* and the nearest neighbor distance *$a_0 = 0.142\ nm$*. Then, top two layers are displaced up vertically to form an ABA trilayer stack with vertical interval *$c_0 = 0.335\ nm$* and then rotated by two angles *$\theta_{12}$* and *$\theta_{23}$*, where *$\theta_{12}$* is the angle between the layer 1 and 2 and $\theta_{23}$ is the angle between layer 2 and 3. It is worth to note that we also performed simulation by rotating an AAA/ABC trilayer instead if an ABA trilayer, and found that the details of the sublattice-locking pattern at very low angles (<~0.5 deg) sensitively depend on the initial stacking-order of a trilayer before twisting; however, the moiré lattices at higher angles are insensitive to whether the initial stacking order is ABA or ABC. After individually rotating the three graphene sheets (**Fig.1 (a)**), we then imposed intralayer and interlayer potentials with approximations based on Brenner [23] and Kolmogorov–Crespi potentials [24], respectively, to represent a HTTG sample.

The molecular dynamics simulation minimizes the system energy by employing the fast inertial relaxation engine (FIRE) algorithm [25], with a termination condition for the change of the system potential energy to be less than 0.5 ppm in an iteration. We normally obtain the fully relaxed ground-state configuration in less than 5000 iteration steps. The MD simulation we employed in this work has advantages over the previous relaxation studies that used calculations based on configurational energy methods [20,21]; for example, the direct simulation of realistic samples with a large size reveals accurate details of the relaxed lattice and the results are straight-forward for an interpretation, thus readily applicable to new complex 2D materials heterostructures. As a trade-off, the computational cost is comparably high; our simulations usually run for 12 hours on an Ubuntu machine with 128 CPU cores and 512 GB ram, where the built-in option of the LAMMPS package allows the simulation to run with 100~200 parallel threads.

**II.2. Tight-binding model Hamiltonian calculation**

To obtain the electronic spectra of lattice-relaxed HTTG at various twisting angles, which don't have a commensurate atomic lattice and a unit cell in general, there are two popular options: continuum-limit calculation and tight-binding model calculation. While the range of validity is questionable at very small angles, a continuum-limit calculation of electronic spectra is an attractive choice with low computational cost and was successfully applied to obtain the electronic density of states (DOS) of twisted bilayer graphene systems [11]. This approach however poses a problem in HTTG where a large multi-scale domain structure is expected [20]. We instead solved a tight-binding Hamiltonian by directly importing the resulting atomic coordinates from the results of LAMMPS molecular dynamics simulations, using open-source *pybinding* package [26] . The tight-binding model calculation is well adapted to account for graphitic layers and has been successfully applied to various multilayer graphene systems to study their exotic electronic spectra [27,28].

To determine the interlayer and intralayer off-diagonal tunneling terms in the Hamiltonian that accounts for the modified coordinates of a relaxed lattice, we accommodate the following approximation [12] with a cutoff distance ($d < 0.6$ mm) to ignore small terms to limit the computational cost:

$$t_{\square} = t_1 exp(-\frac{d-a_0}{0.16a_0})\{1 - (\frac{d_z}{d})^2\} + t_2 exp(-\frac{d-c_0}{0.16a_0})(\frac{d_z}{d})^2$$

, where $d$ is the interatomic distance, $d_z$ is the our-of-plane component of $d$, $c_0$ is the interlayer distance between the graphene layers before relaxation, and $a_0$ is the nearest neighborhood atomic distance in the same graphene layer before relaxation. The intralayer and interlayer tunneling coefficients $t_1$ and $t_2$ are set to be $-2.9$ eV and $-0.37$ eV, respectively. The use of the cutoff distance helps to reduce the calculation time significantly while the results and conclusion of this study basically remain the same. We use pybinding's kernel polynomial (KPM) approximation method [29] with Jackson kernel and a broadening factor of 15 meV to evaluate the total density of states (TDOS) and local density of states (LDOS).

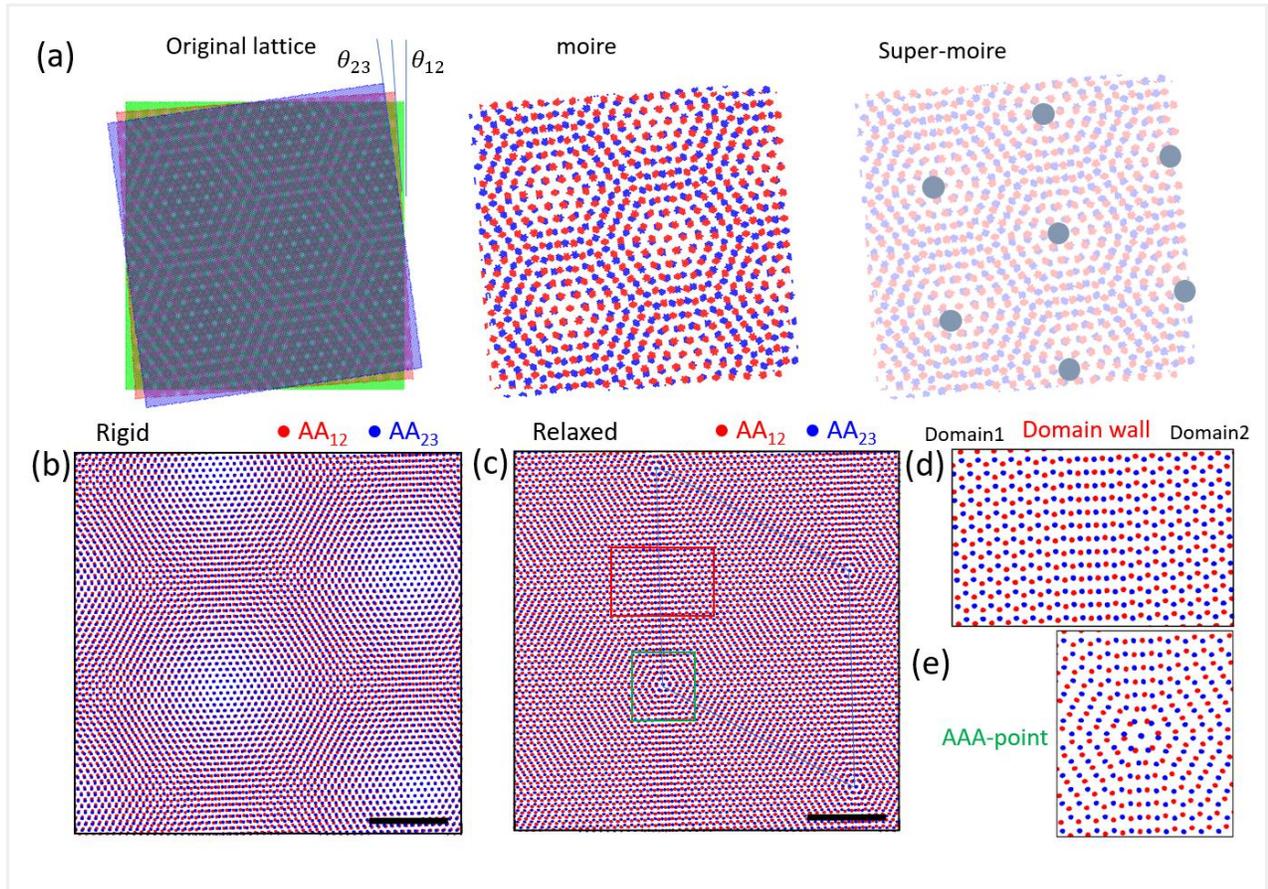

FIG 1. **Helically twisted trilayer graphene** (a) A schematic picture of a helically twisted trilayer graphene system. The two angles $\theta_{12}$ and $\theta_{23}$ have the same sign. The twisting-induced moiré superlattice and the supermoiré lattice are visualized. The atomic configuration (b) before and (c) after relaxation at $\theta = \theta_{12} = \theta_{23} = 1.8$ deg. The blue parallelogram represents a supermoiré unit cell. (d) The distinct anti-symmetric patterns between two adjacent domains and a stripe pattern in a domain wall are shown in the zoom-in view of the red rectangle in (c). (e) The zoomed-in view of the green rectangle in (c). The most symmetric AAA-point emerges when two moiré AA sites overlap and lead to AAA-sublattice aligned throughout layer 1,2 and 3. A particularly interesting shape appears in the domain wall and the AAA-point. The scale bars in (b) and (c) are both 90 nm.

## III. RESULTS

### III.1. Simulation of lattice relaxation

A helical trilayer graphene (HTTG) with equal twisting angles ($\theta = \theta_{12} = \theta_{23}$) presents an interesting case of two moiré patterns with the equal population, one per each interface of the trilayer, coexisting (see **Fig. 1(a)**). When considering a rigid lattice without relaxation, in **Fig. 1(b)**, the system with $\theta = 1.8$ deg displays a super-moiré (moiré-of-moiré) pattern induced from the spatial interference of two moiré lattices. The super-moiré pattern, originating from a slowly varying displacement between two moiré superlattices, may be regarded as an higher-order version of the original moiré pattern realized in twisted bilayer graphene systems [1,2]. Note that HTTG has three length scales of different order of magnitudes; the atomic length scale $L_a = a_0$, the moiré length scale $L_m = a_0/2 \sin(\theta/2)$ and the super-moiré length scale $L_{sm} = a_0/(2\sin(\theta/2))^2$, where $a_0$ is the graphene lattice constant. Qualitatively similar to the moiré pattern based on atomic lattices, we expect the super-moiré to be prone to relaxation. When the lattice is fully relaxed, the pattern induced by the modified atomic coordinates is evident, in **Fig. 1(c)**, where the $AA_{12}$ (AA sublattice alignment between layer 1 and 2 denoted by red dots) and $AA_{23}$ (AA sublattice alignment between layer 2 and 3 denoted by blue dots) stackings form commensurate lattice orders that extend in area and generate domain structures, in qualitative accordance with ref [20]. The size of the domain is ~250 nm and nearly coincides with the supermoiré length scale $L_{sm}$ of the unrelaxed case at $\theta = 1.8$ deg. The commensurate arrangement of two moiré sites inside each domain and the change of the stacking order across the domain wall are clearly visible in the zoomed-in views in **Fig. 1(d)**; deep inside a domain after relaxation, two moiré features - which were incommensurate before the relaxation - align with each other forming a hexagonal commensurate lattice while, in transitioning from the hexagonal arrangement of one the domains to the other in the next domain, the two moiré AA sites seem to align in lines resulting in a stripe pattern. On the other hand, in the region near the point where three domain walls meet (AAA-point, see **Fig. 1(e)**), the moiré lattices are heavily strained with a swirling pattern centered around the point. The shapes of the commensurate domains and the boundaries seem to reflect the surface tension of the two interacting moiré lattices, where the lattice coordinates relax towards more stable configurations. Thus, the comprehensive understanding of how two moirés form a certain shape requires a detailed model that accounts for the moiré lattice rigidity and the interaction of two moirés.

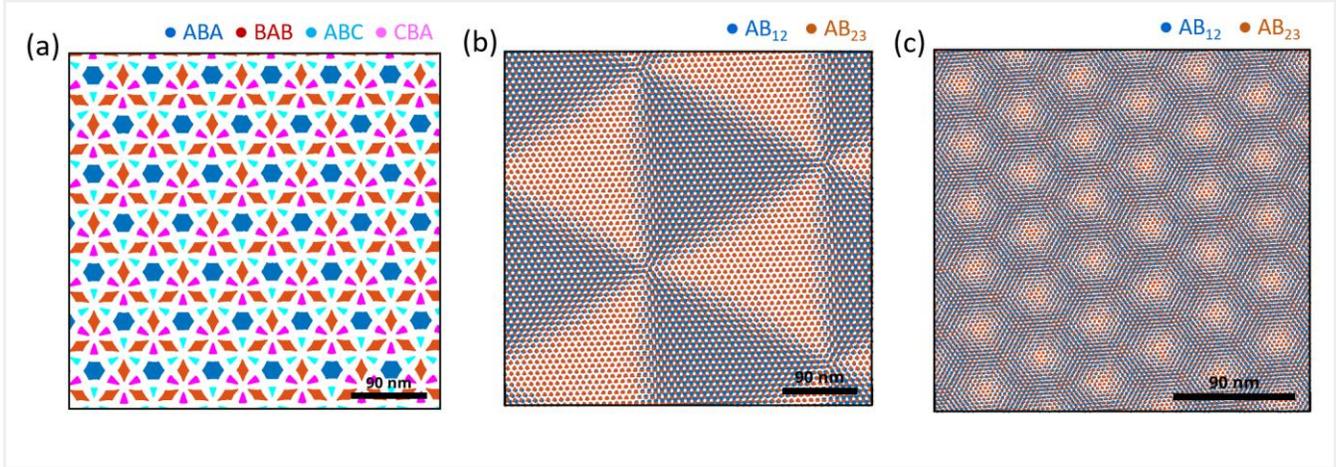

**FIG 2. Angle-dependent relaxation pattern in three different regimes** The locations of interlayer-aligned atomic sublattices are plotted. (a) At $\theta = 0.2$ deg, ABA/BAB/ABC/CBA sublattice-locked areas occupy the majority of the sample. (b-c) At $\theta = 1.8$ deg and 4.0 deg, the area where sublattices of all three layers are aligned is negligible, but there still are AB/BA sublattice-matched locations, as denoted by $AB_{12}$ and $AB_{23}$, between two adjacent layers. The scale bars are all 90 nm.

Taking advantage of the large-scale atomistic calculation, we can directly monitor the system to display an interesting interplay between the atomic lattice and the moiré lattice while varying $\theta$ in a wide range of values. Obviously in the limit of 0 deg rotation (no rotation), the system is known to form a single phase of ABA-stacked or ABC-stacked trilayer graphene [19,30], which by definition has a perfectly-commensurate lattice. At non-zero but very small twisting angles, the lattice relaxation prevents the formation of a moiré pattern but instead leads to local locking of sublattices between layers, which results in domains of ABA-, BAB-, ABC-stacked regions and boundary walls in between, as shown in **Fig. 2 (a)**. This result is consistent with those experimentally observed in a helically twisted trilayer with a very small angle [18,19]. As $\theta$ increases, the size of the domains shrinks and, beyond 0.6 deg, the areas of the sublattice-locked regions become very small, effectively entering the regime of two moirés patterns coexisting – visualized by $AB_{12}$ (AB sublattice alignment between layer 1 and 2) and $AB_{23}$ (AB sublattice alignment between layer 2 and 3) sites in **Fig. 2 (b)** – but twisted by $\theta$. At this angle, however, as already seen in **Fig. 1** the relaxation effect makes the two moirés to form commensurate moiré-lattice domains of an alternating triangular pattern, instead of a super-moiré pattern. Further increasing the angle around $\theta = 4.0$ deg, the moiré commensurate domains also shrink in size and the overall system commensurability diminishes, visibly entering the super-moiré regime (**Fig. 2(c)**).

To gain better insight of the transitions between the commensurate atomic lattices and moiré lattices at different twisting angles, we evaluate the local commensurability factors, $C_R$ and $C_M$, of the atomic lattice and moiré superlattice, respectively. The commensurability of the atomic lattice is fairly straight-forward to evaluate and obtained by counting the number of sublattice matched locations such as ABA/BAB and ABC/CAB sites in the atomic scale and normalized by the whole available sites in a sample. However, quantifying the moiré commensurate lattice needs more consideration. First of all, we start by conventionally defining a moiré lattice with the locations of AA$_{12}$ and the locations of AA$_{23}$ as in **Fig. 1(b)**. However, because the locations of AA sites in principle can not be exactly determined when the moiré lattice and atomic lattice are incommensurate to each other, sets of gaussian probability distribution functions are used to represent the approximate locations of the AA sites. Then, using the distribution functions, we quantify the local commensurability between the two moiré lattice by evaluating the lattice cross-correlation as follows:

$$C_M(X_0, Y_0, \theta) = \frac{\max_{x,y \in S} [\iint_{x',y' \in S} a_{12}(x'-x, y'-y)\, a_{23}(x', y') dx' dy']}{\iint_{x',y' \in S} \frac{1}{2} \{|a_{12}(x', y')|^2 + |a_{23}(x', y')|^2\} dx' dy'}$$

, where $\vec{r} = (x, y)$ are the coordinates defined in-plane of the graphene layers, **S** is a small local region centered at $\vec{r} = (X_0, Y_0) = \vec{R_0}$ with radius set to be 4 times larger than the moiré lattice constant $L_m$, and $a_{12}(\vec{r})$ and $a_{23}(\vec{r})$ are sets of gaussian distribution functions that represent approximate locations of AA$_{12}$ and AA$_{23}$, respectively. Note that the correlation function has the maximum value of 1 when the two moiré lattices are perfectly commensurate in a unit cell. The local commensurability factor of the moiré lattice determined by the function is overlaid as a colored background in **Fig. 3(a)**. When the local factors are averaged within a unit cell, we get the mean commensurability factor $\bar{C}(\theta) = \iint_{u.c.} C(X, Y, \theta)\, dXdY / A_{u.c.}$, where $A_{u.c.}$ is the area of a super-moiré unit cell, such as displayed in **Fig. 1(c)**. In **Fig. 3**, the mean commensurability factors are plotted as a function of twisting angle, where the atomic-lattice commensurability values $\bar{C}_A(\theta)$ rapidly decreases at angle $\theta_A \sim 0.4$ deg while the moiré-lattice commensurability $\bar{C}_M(\theta)$ drops at much larger angle $\theta_M \sim 3.0$ deg. This observation is understood by recognizing the three different length scales at play; the moiré lattice constant $L_m$, the super-moiré lattice constant $L_{sm}$, and the length scale with which the graphene atomic lattices can be deformed. When the angle-dependent lattice constants become larger than the length scale of lattice deformation, the lattice can relax sufficiently to make either atomic or moiré lattices to be commensurate. At the crossover angles of $\theta_A$ and $\theta_M$, we observe that the decrease of atomic commensurability leads to the appearance of a moiré lattice (regime II), and the decrease of the moiré commensurability in turn leads to the super-moiré regime (regime III). The two transition angles thus mark three distinct regimes of a lattice-relaxed helically-stacked trilayer graphene system.

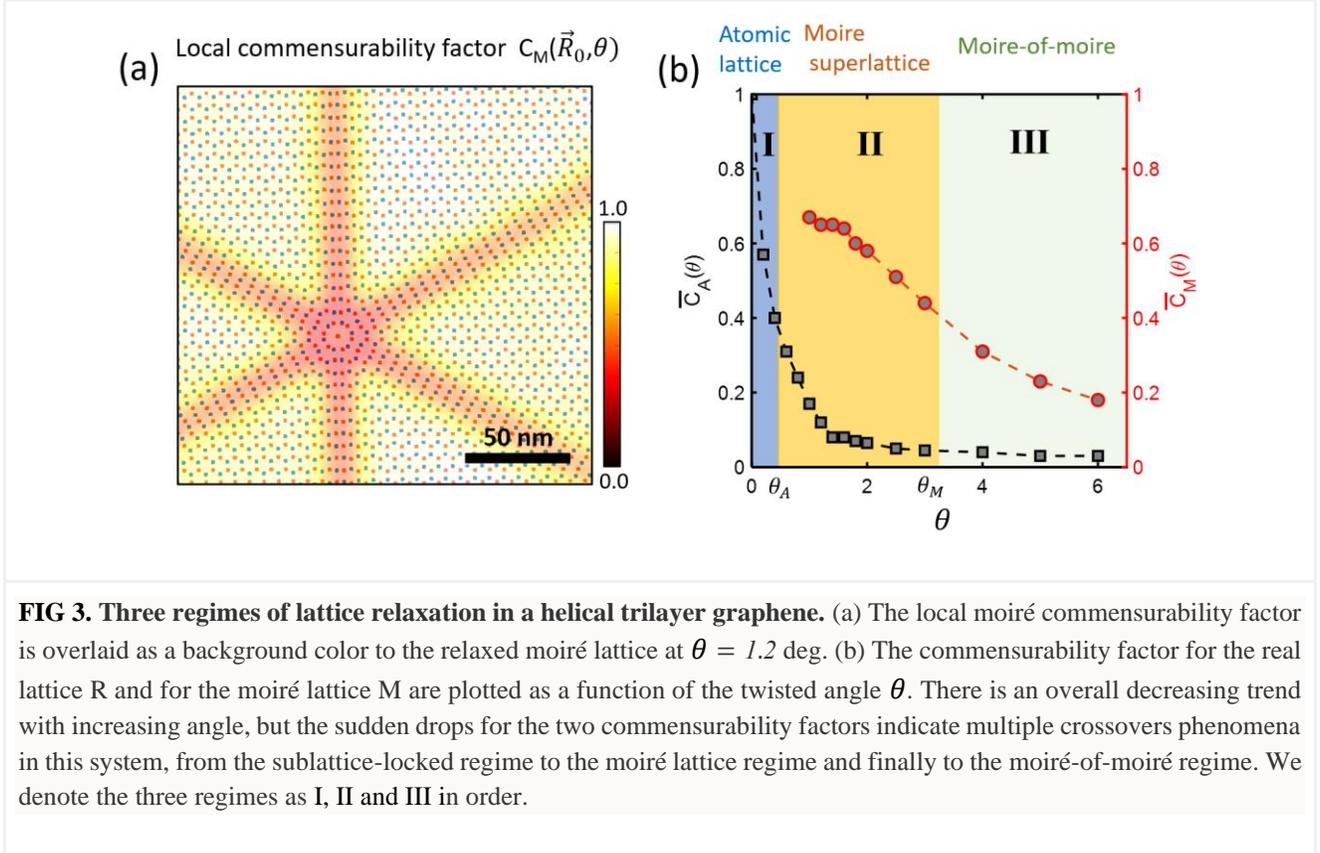

**FIG 3. Three regimes of lattice relaxation in a helical trilayer graphene.** (a) The local moiré commensurability factor is overlaid as a background color to the relaxed moiré lattice at $\theta = 1.2$ deg. (b) The commensurability factor for the real lattice R and for the moiré lattice M are plotted as a function of the twisted angle $\theta$. There is an overall decreasing trend with increasing angle, but the sudden drops for the two commensurability factors indicate multiple crossovers phenomena in this system, from the sublattice-locked regime to the moiré lattice regime and finally to the moiré-of-moiré regime. We denote the three regimes as I, II and III in order.

Among the regimes of lattice-relaxed HTTG, the moiré-commensurate regime II provides an interesting case of realizing a triangular lattice with two moiré bases. The domains of inversion-asymmetric moiré lattices are expected to have non-trivial topological quantum numbers whose values are alternating in sign [12], and these domains are separated by the narrow domain boundary regions that form a macroscopic triangular network as shown above. Since both domain and domain boundary regions likely contribute to the electronic property of a sample as a whole, a proper interpretation of experimental data of micron-sized samples can be challenging and thus a quantitative analysis of the electrical properties of the boundary regions and how they affect the properties of the whole sample would be very informative. In the following section, we focus on the electronic spectra of fully lattice-relaxed HTTG including its peculiar microscopic spectral features.

### III.2. Tight-binding calculation: Electronic density of states

To investigate the effect of lattice relaxation on the electronic degrees of freedom, we performed a tight-binding based calculation that solves the electronic Hamiltonian of a HTTG system, and obtained density of states (DOS) as shown in **Fig. 4**. The tight-binding Hamiltonian is formulated directly from the information of the atomic coordinates obtained by the lattice relaxation simulation at various twisted angles of $\theta = \theta_{12} = \theta_{23}$. As can be seen in angle dependent TDOSs of **Fig. 4(b)**, at small $\theta$, the effect of twisting is concentrated near the energy scale of a few hundred meV around the zero energy (CNP), similar to the twisted bilayer graphene; the characteristic peak near CNP corresponds to the isolated flat band separated by sizable energy gaps (~ 100 meV) from other bands at higher energy. As the twisting angle increases, the energy gap moves to higher energy but the gap remains similar in size while the central peak of the isolated bands becomes broader and breaks into two peaks. This is consistent with the picture that the bandwidth of the band is increasing at high angles. To better identify the effects of lattice relaxation, TDOS before and after the relaxation at $\theta = 1.8$ deg, in **Fig. 4(c)**, are compared to show that the effect of lattice relaxation is dramatic; while the DOS for the relaxed system shows large energy gaps at $E = \pm 80$ meV, they are nearly absent in the rigid system, strongly suggesting that the effects of the lattice relaxation and domain formation played an important role in modifying the low-energy electronic spectra.

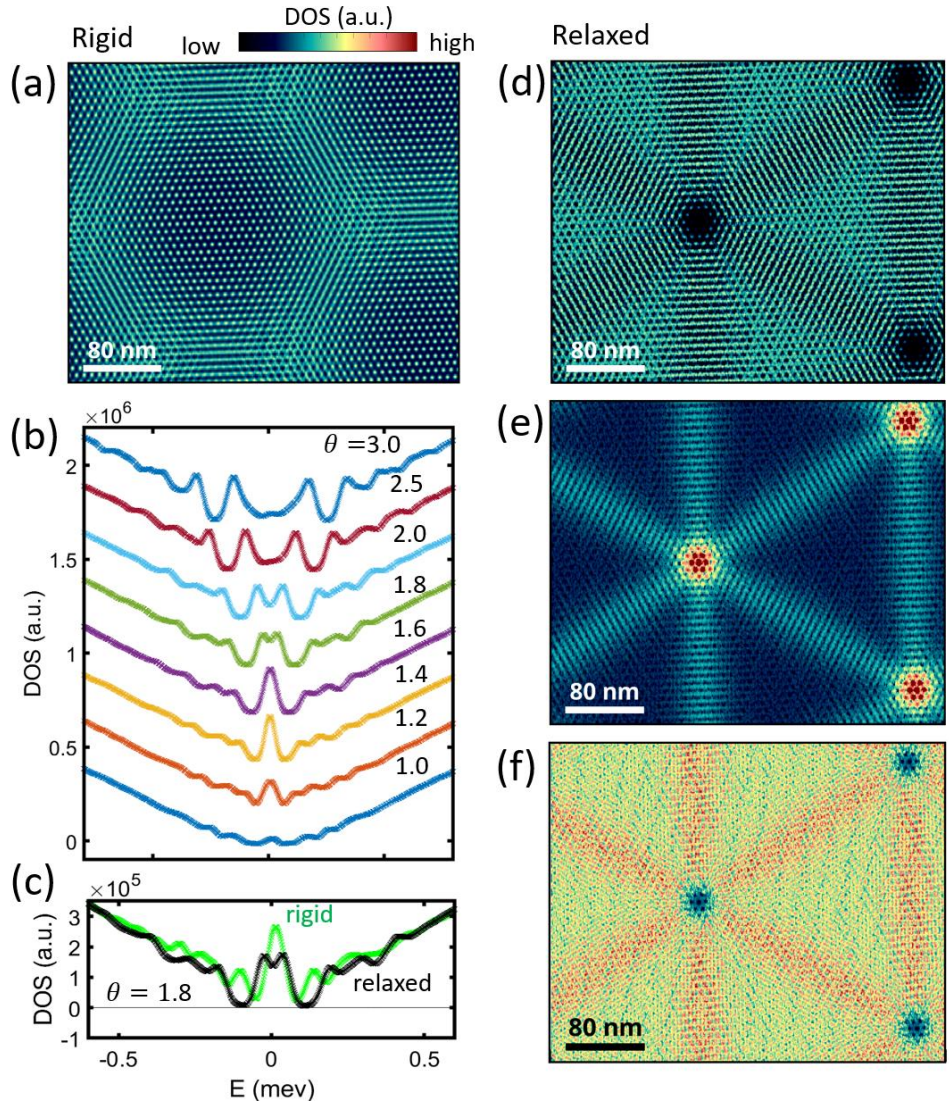

**FIG 4. LDOS evaluated at $\theta = 1.8$ deg.** Local density of states at $\theta = 1.8$ deg (a) before at E = 0 meV and (d,e,f) after relaxation at $E = 0, 80, 225$ meV, respectively. (b) TDOS of a system before (green) and after (black) relaxation at $\theta = 1.8$ deg. (c) TDOS for a relaxed system at various twisting angles.

For further insight on the modified electronic spectra after relaxation, we obtained the local density of states (LDOS) of the system. In **Fig. 4(a) and (d),** the spectra before and after relaxation are evaluated respectively at $E_F = 0$ meV and $\theta = 1.8$ deg, an angle that falls well within the moiré commensurate lattice regime II, as discussed in Section 3. In the case of the electronic spectrum without lattice relaxation, LDOS displays two moiré patterns that overlap in a simple manner with a relative twisting angle, a pattern that is expected from the moiré-of-moiré system. For the case of a relaxed lattice, however, the spectrum is qualitatively different and the formation of spectral features that correspond to the results of the lattice simulation (see also **Fig. 1(c)**) are evident. Including distinct patterns that must be originating from the stacking orders of moiré lattices, a triangular network of the domain boundaries is very noticeable and suggestive of its origin as the domain walls observed in the lattice relaxation simulation. The spatial distribution of the electronic spectra at this low energy bias clearly distinguishes domains with high DOS and domain walls and the high symmetric AAA-points with suppressed DOS. Furthermore, there are peculiar 3-fold weaved patterns inside domains reflecting the unique hexagonal moiré-stacking orders in the region. Note that the stripe patterns on the domain walls appear to break three-fold symmetry locally but there are three domain walls surrounding a domain so that the three-fold symmetry is globally recovered as anticipated.

The LDOS images of the relaxed lattice taken at higher energy values in **Fig. 4 (e)** and **(f)** show that local spectral features are strongly energy dependent. In particular, the spectra of the region inside domains show dramatically different energy dependent behaviors from those of boundary walls; for instance, at $E = 80$ meV, the spectra on domain walls has substantial DOS while the spectra inside domains have a very small number of states due to the dominant energy gap as mentioned with TDOS. We attribute this observation to the dispersing topological mid-gap edge states at the boundary between two topologically distinct domains [20,21]. The remnant states on the domain walls, protected by a topological gap closing, thus give rise to the triangular network of electrically conducting 1D channels. When evaluated at higher energies of ~225 meV, however, the difference of the LDOS between the areas are much less noticeable, highlighting the fact that the moiré lattice stacking *selectively* modifies the low-energy electronic spectra, and the spectra on domain walls and those inside domains behave qualitatively differently as a function of energy owing to different moiré stackings. We like to point out that this would provide an opportunity towards robust and controllable formation of topologically nontrivial macroscopic regions with distinct electronic properties.

Also interestingly, we notice spectral features that belong to a unique lattice configuration of the AAA-points (see also **Fig. 1(e)**). In **Fig. 5(a)**, the spectra calculated at $\theta = 4.0$ deg and $E = 380$ meV shows a pattern of sites with high DOS and the locations of these sites coincide with the lattice regions where $AA_{12}$ and $AA_{23}$ moiré sites overlap, i.e. AAA-points. Such an overlap means that the local atomic lattice is stacked with AAA (and equivalently BBB) sublattice alignment in all three graphene layers, whose configuration has a very high energy cost [31]. They however don't disappear as $\theta$ changes and their existence seems to be protected by the global non-zero twisted angle. As can be seen by comparing **Fig. 4(d)** and **(e),** the electronic spectra of AAA-points are suppressed near $E = 0$ meV but instead the missing states are pushed up to higher energy levels. Even at $\theta = 4.0$ deg, where the moiré lattice is no longer commensurate forming the moiré-of-moiré phase of regime III, the highly energetic electronic states near AAA-points don't disappear but pack more closely to form a pattern of regularly spaced islands with ~50 nm spacing. We expect that these patterns would be measurable using scanning probe techniques and nanoscale angle-resolved spectroscopic probes. It is important to note that the lattice also relaxes along the out-of-plane (vertical) direction displaying sizable corrugations that depend on the local moiré and atomic configurations. As shown by the vertical coordinates of the top and bottom layers around an AAA-point, in **Fig. 5(b)** and **(c)**, the top layer relaxes downwards while the bottom layer shifts upwards as much as 0.02 nm, thus leading to significantly shorter interlayer distance and stronger interlayer tunneling strength. The lattice relaxation along the vertical direction thus provides critical information and needs to be accounted for to comprehensively understand electronic spectra observed in this system.

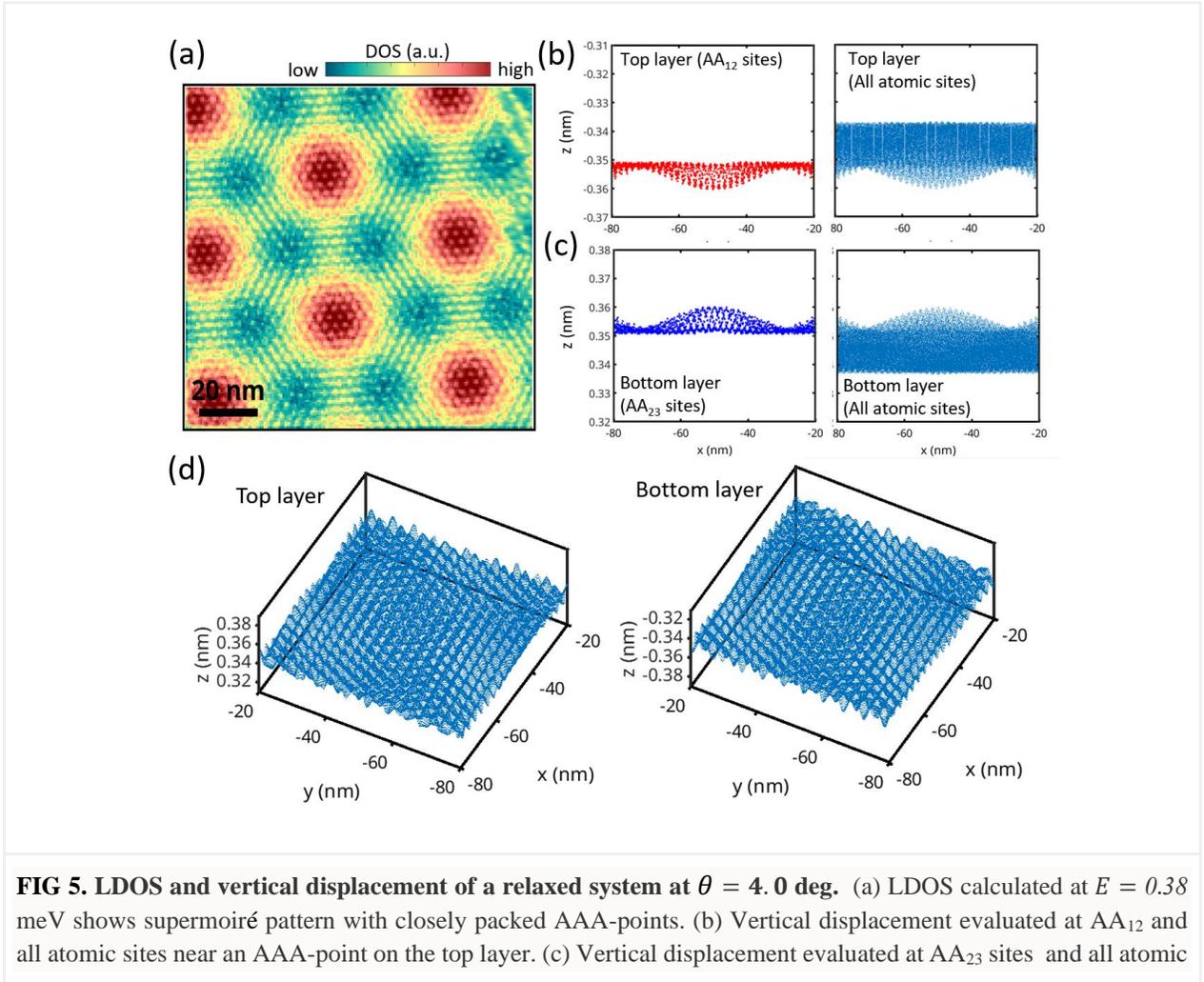

**FIG 5. LDOS and vertical displacement of a relaxed system at $\theta = 4.0$ deg.** (a) LDOS calculated at $E = 0.38$ meV shows supermoiré pattern with closely packed AAA-points. (b) Vertical displacement evaluated at $AA_{12}$ and all atomic sites near an AAA-point on the top layer. (c) Vertical displacement evaluated at $AA_{23}$ sites and all atomic

sites near the same AAA-point on the bottom layer. (d) Perspective views of the atomic coordinates near an AAA-point on the top and bottom layers. The small ripples are layer corrugations at moiré sites.

## IV. DISCUSSION

As the moiré commensurate orders and incommensurate supermoiré features can realize in a macroscopic scale, the distinct spectral signatures would be measurable with already available tools, such as scanning tunneling microscopy (STM) by exposing the top surface of HTTG without a hBN dielectric [32]. But more practically, spatial structures in electron spectra can also be studied even if the sample is embedded in a dielectric material, using techniques that are sensitive to local thermodynamic electronic response from the sample. In particular, the spatial features of the LDOS of the relaxed structure would be reflected in the electronic compressibility and local capacitance, to which scanning capacitance microscopy or scanning single-electron transistors (scanning SET) are sensitive [33]. The scale of the macroscopic domains that exist over the whole area in a sample is an order of 100 nm or larger for twisting angles below 3 deg, thus a scanning thermodynamic tool such as Kelvin probe microscope is also one of good candidates to detect the unique domain structures and study the thermodynamic and transport properties of a sample with the macroscopic domains and a domain wall network.

The intricate interplay of the atomic lattice and the moiré lattice is the underlying mechanism that leads to a significant modification of the electronic degrees of freedom, based on which unique spectral features of HTTG are realized. In that sense, sets of local spectral and thermodynamic information as a function of twisting angle would be very instrumental to fully understand the role of the lattice relaxation played on electronic degrees of freedom in this system and twisted multilayer systems in general. For example, compared to those observed in other twisted graphene systems, significant lattice relaxation occurs near the magic-angle - estimated to be around 1.5 ~ 1.8 deg [20] (see also **Fig 4(c)**), considerably higher than those observed in other twisted systems. The effect of lattice relaxation on the bandwidth of the isolated band near CNP and a potential modification of the magic-angle using lattice engineering, however, are yet to be fully investigated. Our analysis of the electronic spectra suggest that strain engineering to rearrange moiré lattices in this system would be one of the ways to qualitatively and significantly modify the electronic bands at low energy and may lead to emergent quantum phases.

## V. CONCLUSION AND OUTLOOK

In conclusion, we systematically investigated lattice relaxation of helically twisted trilayer graphenes, and observed cascades of lattice relaxations as twisting angle changes. The emergence of triangular moiré-domain structure observed in the lattice simulation and subsequent tight-binding calculations based on the modified lattice allowed us to identify distinct patterns of low-energy electronic spectra that are induced by the lattice relaxation. We suggest that the spectral patterns associated with unique moiré stackings can be visualized with scanning probe tools such as scanning SET and STM. Our work demonstrates that the electronic and lattice degrees of freedom have intimate relationships in HTTG, and quantitative estimation of lattice deformation is critical for studies of electronic phases in Van der Waals multilayer systems.

## VI. ACKNOWLEDGEMENT

J.J acknowledges fruitful discussions on the lattice commensurability with Seungwon Jung and technical support on the Ubuntu computational server from Youngoh Son. This work was supported by the National Research Foundation of Korea grants funded by the Ministry of Science and ICT (Grant Nos. 2019R1C1C1006520, 2020R1A5A1016518, RS-2023-00258359), the Institute for Basic Science of Korea (Grant No. IBS-R009- D1), the Core Center Program (2021R1A6C101B418) by the Ministry of Education of Korea, Creative-Pioneering Researcher Program through Seoul National University and Samsung DS Basic Research Program (Project No. 0409-20230298).

## VI. STATEMENTS AND DECLARATIONS

Competing interests

The authors have no other competing interests to declare that are relevant to the content of this article.

Data availability

Data used to support the result of this study are available from the corresponding author upon a reasonable request.